\documentclass[useAMS,usenatbib]{mn2e}
\usepackage{epsfig}
\usepackage{amssymb}
\usepackage{amsmath}
\usepackage{graphicx}
\usepackage{booktabs}
\usepackage{float}
\usepackage{url}
\usepackage{color}

\def\Msun{\ifmmode{M_\odot}\else$M_\odot$\fi}
\def\msun{\ifmmode{M_\odot}\else$M_\odot$\fi}

\newcommand{\lsimeq}{\mbox{$\, \stackrel{\scriptstyle <}{\scriptstyle
\sim}\,$}}

\title[Merging Binary Stars and the magnetic white dwarfs]
{Merging binaries and magnetic white dwarfs}

\author[Briggs, Ferrario, Tout, Wickramasinghe \& Hurley]
{Gordon P.~Briggs$^1$, Lilia Ferrario$^1$, Christopher A.~Tout$^{1,2,3}$,
\newauthor Dayal T.~Wickramasinghe$^1$ and Jarrod R.~Hurley$^4$\\
$^1$Mathematical Sciences Institute, The Australian National University, ACT 0200, Australia\\
$^2$Institute of Astronomy, The Observatories, Madingley Road,
Cambridge CB3 0HA, UK\\
$^3$Monash Centre for Astrophysics (MoCA), School of Mathematical Sciences, Monash University, VIC 3800, Australia\\
$^4$Centre for Astrophysics \& Supercomputing, Swinburne University of Technology, Hawthorn, VIC 3122, Australia
}
 
\begin{document}

\date{Accepted.  Received ; in original form} 
\pagerange{\pageref{firstpage}--\pageref{lastpage}} \pubyear{}

\maketitle

\label{firstpage}

\begin{abstract}
  A magnetic dynamo driven by differential rotation generated when
  stars merge can explain strong fields in certain classes of magnetic
  stars, including the high field magnetic white dwarfs (HFMWDs).  In
  their case the site of the differential rotation has been variously
  proposed to be within a common envelope, the massive hot outer
  regions of a merged degenerate core or an accretion disc formed by a
  tidally disrupted companion that is subsequently incorporated into a
  degenerate core.  We synthesize a population of binary systems to
  investigate the stellar merging hypothesis for observed single
  HFMWDs.  Our calculations provide mass distribution and the
  fractions of white dwarfs that merge during a common envelope phase
  or as double degenerate systems in a post common envelope phase.  We
  vary the common envelope efficiency parameter $\alpha$ and compare
  with observations.  We find that this hypothesis can explain both
  the observed incidence of magnetism and the mass distribution of
  HFMWDs for a wide range of $\alpha$.  In this model, the majority of
  the HFMWDs are of the Carbon Oxygen type and merge within a common
  envelope.  Less than about a quarter of a per cent of HFMWDs
  originate from double degenerate stars that merge after common
  envelope evolution and these populate the high-mass tail of the
  HFMWD mass distribution.

\end{abstract}


\section{Introduction}
\label{Introd}
Magnetic fields are seen in main-sequence stars of most spectral types.
They are usually considered to be either of fossil origin, arising from a
conserved primordial field, or generated in a contemporary dynamo
\citep{Mestel2005}.  The latter is the accepted explanation for magnetic
stars with convective envelopes such as the low-mass ($M < 1.5\,$\msun)
main-sequence stars.  The origin of the fields in the higher-mass magnetic
Ap and Bp main-sequence stars with radiative envelopes is less certain.
While a fossil origin remains possible, it has been proposed that
magnetic fields may be generated by a dynamo mechanism driven by
various instabilities, including the magnetorotational instability, in
differentially rotating radiative regions of single stars 
\citep[see e.g.][]{potter2012}.

The origin of the high field magnetic white dwarfs (HFMWDs) has been
the topic of much discussion in recent years.  The incidence of
magnetism in white dwarfs in the high field group ($B>10^6$\,G) is
estimated to be about 8-16\,per cent \citep{Liebert2003, Kawka2007}.
A traditional explanation has been that the fields are of a fossil
origin from the main sequence with magnetic flux conserved in some way
during evolution to the white dwarf phase \citep{Mestel2005}.
\citet{Kawka2007} pointed out that the strongly magnetic Ap and
Bp~stars could not be their sole progenitors because the birth rate of
these main-sequence stars is insufficient to explain the observed
birth rate of the HFMWDs.  However this turned out not to be a strong
argument against the fossil hypothesis.  In an earlier paper
\citet{wickramasinghe2005} noted that it could be reconciled if about
$40$\,per cent of late B~stars had fields below the observed threshold
for Ap and Bp~stars.  This would be consistent with the observations
of \citet{power2008} who conducted a volume-limited study of the
magnetic Ap and Bp~stars within 100\,pc of the Sun.  Their study has
shown that the incidence of magnetism in intermediate mass stars
increases with the mass of the stars.  At $1.7\,\rm\msun$ the fraction
of magnetic among non-magnetic stars is only $0.1$\,per cent, while at
$3.5\,\rm\msun$ it is 37.5\,per cent.

Some 50\,per cent of stars are in binary systems. As these evolve some
can interact and merge. So we may expect that some stars that appear
single today are the result of the merging of two stars.  The
possibility of generating strong magnetic fields during such merging
events has often been discussed in the literature as an alternative
explanation for magnetic fields in certain classes of stellar object.
Indeed, as an alternative to the fossil field model,
\citet{ferrario2009} proposed that the strong fields in the magnetic
A, B and O~stars are generated as stars merge.

Here we focus on the hypothesis that the entire class of
HFMWDs with fields $10^6<B/{\rm G}<10^9$ owe their magnetic 
fields to merging \citep{tout2008}.  This model was
first devised to explain the observation that there are no examples
of HFMWDs in wide binary systems with late-type companions while
a high fraction of non-magnetic white dwarfs are found in such systems
\citep{liebert2005}. 

In the common envelope scenario, when a giant star fills its Roche
lobe, unstable mass transfer can lead to a state in which the giant's
envelope engulfs both cores.  As the two cores spiral together, energy
and angular momentum are transferred from their orbit to the
differentially rotating common envelope until it is ejected, leaving
behind a close binary system, or a merged single object.  In the original
model for formation of HFMWDs \citet{tout2008} envisaged that the
fields are generated by a dynamo in the common envelope and diffuse
into the partially degenerate outer layers of the proto-white dwarf
before the common envelope is ejected.  If the end product is a single
star it can have a highly magnetic core and if it is a very close
binary, it can become a magnetic cataclysmic variable.
\citet{potter2010} attempted to model this phenomenon and found a
potential problem in that the time-scale for the diffusion of the
field into the white dwarf is generally significantly longer than the
expected common envelope lifetime.

\citet{wickramasinghe2014} suggested that strong magnetic fields in white dwarfs
are generated by a dynamo process that feeds on the differential
rotation in the merged object as it forms.  A weak poloidal seed field
that is already present in the pre-white dwarf core is amplified by the dynamo
to a strong field that is independent of its initial strength but
depends on the amount of the initial differential rotation.  We note
in this context that weak fields of $B\le1\,$kG may be present in
most white dwarfs \citep{landstreet2012}.  Presumably these can be
generated in a core--envelope dynamo in the normal course of stellar
evolution.

\citet{Nordhaus2011} proposed an alternative but similar model
(hereinafter the disc field model).  They noted that if the companion
were of sufficiently low mass it would be disrupted while merging
and form a massive accretion disc around the proto-white dwarf.  Fields
generated in the disc via the magnetorotational instability or other
hydrodynamical instabilities could then be advected on to the surface
of the proto-white dwarf and so form a HFMWD.  Such a model could apply to some
merging cores within the common envelope, depending on component masses, and to
post-common envelope merging double degenerate systems (DDs).  It depends on the
time-scale for the diffusion of the field into the white dwarf
envelope.

\citet{garcia2012} used the results of a three-dimensional
hydrodynamic simulation of merging DDs to argue that a massive hot and
differentially rotating convective corona forms around the more
massive component and used equipartition arguments to estimate that
fields of about $3\times 10^{10}\,$G could be generated.  They also
presented a population synthesis study of white dwarfs that formed
specifically as merging DDs, assuming a common envelope energy
efficiency parameter $\alpha=0.25$, and showed that there is general
agreement with the observed properties of high-mass white dwarfs
($M_{\rm WD}>0.8\,\rm{\Msun}$) and HFMWDs.  However
they did not consider merging when the companion is a
non-degenerate star.

We hypothesize that single white dwarfs that demonstrate a strong
magnetic field are the result of merging events, so we carry out a
comprehensive population synthesis study of merging binary systems for
different common envelope efficiencies $\alpha$.  We consider all
possible routes that could lead to a single white dwarf.  We isolate
the white dwarfs formed by the merging of two degenerate cores, either
as white dwarfs, a red giant plus a white dwarf or two red giants,
from those formed by a giant merging with a main-sequence star and
show that the observed properties of the HFMWDs are generally
consistent with the common envelope hypothesis for $0.1 \le \alpha \le
0.3$.  Both groups contribute to the observed distribution but
main-sequence companions merging with degenerate cores of giants form
most of the HFMWDs.

\section{Common Envelope Evolution and Formulism}

When one of the stars in a binary system becomes a giant, it expands
and overfills its Roche lobe. Mass transfer soon proceeds typically,
but not always, on a dynamical time-scale \citep{Han2002}.  The giant
envelope rapidly engulfs both the companion star and the core of the
donor to form a common envelope.  The two dense cores, that of the
giant and the accreting star itself, interact with the envelope,
transferring to it orbital energy and angular momentum.  The envelope
can be partly or wholly ejected and the orbit of the engulfed star
shrinks.  It is not known how long this process takes but it is
generally thought to last for more of a dynamical stellar time-scale
than a thermal or nuclear time-scale.  It probably has never been
observed.  If the companion succeeds in fully ejecting the envelope
the two cores survive in a binary system with a much smaller
separation.  If the envelope is not fully ejected the orbit may
completely decays and the two stars coalesce.  When the envelope of a
giant engulfs a degenerate companion the two cores can merge but if
the companion is non-degenerate it either merges with the envelope or
accretes on to the giant core.  When the initial masses of the two
stars are within a few percent both can expand to giants at the same
time and Roche lobe overflow (RLOF) leads to a double common envelope.

The common envelope process was first proposed to explain binary star systems,
such as cataclysmic variables (CVs), whose orbital separations are
smaller than the original radius of the progenitor primary star.  A
mechanism was needed to explain how this could occur.  The possible
existence of common envelopes was first proposed by
\citet{Bisnovatyi1971}.  Its qualitative description is based on
evolutionary necessity rather than mathematical physics.  While it is
sufficient to explain a variety of exotic stars and binaries that
could not otherwise be explained, a full mathematical model has
yet to be developed to describe the interaction in detail and to test
the various theories.

A simple quantitative model of common envelope evolution is the energy or $\alpha$
formulism \citep{vandenHeuvel1976}.  For this the change
in orbital energy $\Delta E\rm_{orb}$ of the in-spiralling cores
is equated to the energy required to eject the envelope to infinity,
the binding energy $E\rm{_{bind}}$.  The total orbital energy,
kinetic plus potential, of a binary star with masses ${m_1}$
and~${m_2}$ and separation $a$ is $E_{\rm_{orb}} =-{\rm G}m_{\rm 1}m_{\rm
  2}/2a$.  However the envelope ejection cannot be completely
efficient so \citet{Livio1988} introduced an efficiency parameter
$\alpha$ to allow for the fraction of the orbital energy
actually used to eject the envelope.
\begin{equation}
\Delta E_\textrm{orb} = \alpha E_{\rm{bind}}.
\end{equation}
 Following \citet{Tauris2001} we use a form of the binding energy
 that depends on the detailed structure of the giant envelope and adopt 
\begin{equation}
E_{\rm bind} = -\frac{{\rm{G}}m_{\rm 1} m_{\rm{1 ,env}}}{{\lambda}R_{1}},
\end{equation}
where {\it{R}}$_{\rm1}$ is the radius of the primary envelope.  The
constant $\lambda$ was introduced by \citet{deKool1990} to
characterize the envelope structure.  Our $\lambda$ depends on the
structure of the particular star under consideration.  It is sensitive
to how the inner boundary between the envelope and the remnant core is
identified \citep{Tauris2001} and includes the contributions from the
thermal energy of the envelope on the assumption that it remains in
equilibrium as it is ejected.

The initial orbital energy is that of the secondary star $m_2$ and
the primary core $m_{\rm 1,c}$ at the orbital separation $a_{\rm i}$
at the beginning of common envelope and is given by
\begin{equation}
E_{\rm orb, i}=-\frac12\frac{{\rm G}m_{\rm 1,c} m_{\rm 2}}{a_{\rm i}}
\end{equation}
and the final orbital energy is
\begin{equation}\label{Ermf}
E_{\rm orb, f}=-\frac12\frac{{\rm G}m_{\rm 1,c} m_{2}}{a_{\rm f}},
\end{equation}
where $a_{\rm f}$ is the final orbital separation.  Thus we have
\begin{equation}\label{deltaEorb}
\Delta E_{\rm orb}=E_{\rm orb, f}-E_{\rm orb, i}.
\end{equation}
From this we can calculate $a_{\rm f}$ which is the separation of the
new binary if the cores do not merge.  If $a_{\rm f}$ is so small that
either core would overfill its new Roche lobe, then the cores are
considered to merge when $a_{\rm f}$ is such that the core just fills
its Roche lobe. Setting $a_f$ to this separation we calculate $E_{\rm
  orb,f}$ and $\Delta E_{\rm orb}$ with equations \ref{Ermf} and
\ref{deltaEorb}. Then we calculate a final binding energy for the
envelope around the merged core
\begin{equation}
E_{\rm bind,f}=E_{\rm bind, i}+\frac{\Delta E_{\rm orb}}{\alpha}.
\end{equation}
Assuming this envelope has a normal giant structure $R(m,m_{\rm c})$
we calculate how much mass must be lost.  In the case of a double
common envelope, the initial orbital energy is that of both cores and
the binding energies of the two envelopes added.

Some difficulties with the energy formulation arise because $\alpha$
can depend on the duration of the common envelope phase.  If it lasts
longer than a nuclear or thermal time-scale then alterations in the
envelope, owing to adjustments in its thermal equilibrium, can change
its structure and hence $\lambda$.  Changes to the energy output from
the core, owing to the decreasing weight of the diminishing envelope,
can also affect the thermal equilibrium and thence $\lambda$.  We do
not consider these complications in this work. Nor do we include
ionization and dissociation energy, as proposed by \citet{Han1994} in
the envelope binding energy.

\section{Population synthesis calculations}
\label{sec:calculations}

\begin{table}
\centering
\caption{Stellar types distinguished within the {\sc bse}
  algorithms.  \label{tab:bsetypes}}
\vspace{5pt}
\begin{tabular}{ r l }
\hline
Type & Description\\
\hline
  0.& Deep or fully convective low-mass MS star (CS)\\
  1.& Main-sequence star (MS) \\
  2.& Hertzsprung gap star (HG) \\
  3.& First giant branch (RGB) \\
  4.& Core helium Burning \\
  5.& First asymptotic giant branch (early AGB) \\
  6.& Second asymptotic giant branch (late AGB) \\
  7.& Main-sequence naked helium star \\
  8.& Hertzsprung gap naked helium star \\
  9.& Giant branch naked helium star \\
10.& Helium white dwarf (He white dwarf)\\
11.& Carbon/oxygen white dwarf (CO white dwarf)\\
12.& Oxygen/neon white dwarf (ONe white dwarf)\\
13.& Neutron star \\
14.& Black hole \\
15.& Massless supernova/remnant \\
\hline
\end{tabular}
\end{table}

We evolve synthetic populations of binary star systems from the
zero-age main sequence (ZAMS).  Each system requires three initial
parameters, the primary star mass, the secondary star mass and the
orbital period.  The primary masses $M_1$ are allocated between
$0.8$ and $12.0\rm{\,\msun}$ and the secondary star masses
$M_2$ between $0.1$ and $12.0\,\rm{\msun}$.  The binary
orbits are specified by a period $P_0$ at ZAMS between $0.1$ and
$10\,000\,$d and zero eccentricity.  Each parameter was uniformly
sampled on a logarithmic scale for 200\, divisions.  This scheme gives
a synthetic population of some 6 million binary systems.  We calculate
the effective number of actual binary systems by assuming that the
primary stars are distributed according to Salpeter's mass function
\citep{Salpeter1955} $N(M)\,{\rm d}M\propto M^{-2.35}\,{\rm d}M$,
where $N(M)\,{\rm d}M$ is the number of stars with masses between $M$
and $M+{\rm d}M$, and that the secondary stars follow a flat mass
ratio distribution for $q\le 1$ \citep[e.g.][]{Ferrario2012}.  The initial
period distribution was taken to be logarithmically uniform in the
range $-1\le \log_{10} P_0/{\rm d} \le 4$.

\begin{table*}
  \caption{Fraction of binary systems that merge during common
    envelope for various values of 
    $\alpha$.  The fraction of white dwarfs born from merged stars in a single
    generation of binary systems of age 9.5\,Gyr (the age of the Galactic disc) is $N$.
    The remaining six  columns give the smallest and the largest
    parameters on the search grid for systems that are found to have
    merged.  The parameters are the progenitors' ZAMS masses and orbital period.}
\centering
\begin{tabular}{ c c c c r c c r}
\hline
$\alpha$ & $N\,$per cent & $M_{1_{\rm min}}/\msun$ & $M_{2_{\rm
    min}}/\msun$ & $P_{0_{\rm min}}/$d &$M_{1_{\rm max}}/\msun
$&$M_{2_{\rm max}}/\msun $ & $P_{0_{\rm max}}/$d \\
\hline
0.05   & 11.58   &  1.08  &  0.10  &   348.9  & 11.06  &  2.77  &     16.3 \\ 
0.10   & 10.35   &  1.08  &  0.10  &   195.6  & 11.06  &  2.90  &     20.5 \\
0.20   &   8.86   &  1.08  &  0.10  &     97.7  & 11.21  &  2.77  &     20.5 \\
0.25   &   8.17   &  1.08  &  0.10  &     82.1  & 11.21  &  4.06  &   932.9 \\
0.30   &   7.55   &  1.08  &  0.10  &     65.2  & 11.21  &  4.06  &   784.3 \\
0.40   &   6.51   &  1.08  &  0.10  &     48.8  & 11.21  &  4.06  &   587.3 \\
0.50   &   5.70   &  1.08  &  0.10  &     38.7  & 11.21  &  4.06  &   493.7 \\
0.60   &   5.06   &  1.08  &  0.10  &     30.7  & 11.21  &  4.06  &   391.7 \\
0.70   &   4.60   &  1.08  &  0.10  &     25.8  & 11.21  &  3.87  &   195.6 \\
0.80   &   4.18   &  1.08  &  0.10  &     23.7  & 11.21  &  3.12  &     82.1 \\
0.90   &   3.75   &  1.08  &  0.10  &     19.3  & 11.06  &  4.06  &   415.0 \\
\hline
\end{tabular}
\label{tab:statsCE}
\end{table*}

\begin{table*}
  \caption{Fraction of merging DD systems, white dwarfs formed by merging of
    two degenerate objects outside a common envelope in a single generation of binary
    systems of age 9.5\,Gyr.  Other columns are as in Table~\ref{tab:statsCE}  }
   \centering
  \begin{tabular}{ c c c c r c c r }
\hline
$\alpha$ & $N\,$per cent & $M_{1_{\rm min}}/\msun$ & $M_{2_{\rm
      min}}/\msun$ & $P_{0_{\rm min}}/$d & $M_{1_{\rm max}}/\msun $ &
  $M_{2_{\rm max}}/\msun$ & $P_{0_{\rm max}}/$d \\
\hline
0.05   & 4.49 x 10$^{-5}$   &  2.41  &  1.79  & 1867.9  &   4.21  &  2.17  &  3331.3  \\
0.10   & 4.89 x 10$^{-4}$   &  2.02  &  1.79  & 1245.9  &  4.21  &  2.28  &  2097.0  \\
0.20   & 1.01 x 10$^{-4}$   &  1.99  &  1.98  &   932.9  &  4.21  &  2.28  &  1867.9  \\
0.25   & 1.29 x 10$^{-4}$   &  1.99  &  1.98  &   784.3  &  4.21  &  2.28  &  1867.9  \\
0.30   & 1.69 x 10$^{-4}$   &  1.52  &  1.52  &   587.3  &  4.27  &  2.23  &  1867.9  \\
0.40   & 2.62 x 10$^{-4}$   &  1.52  &  1.52  &   587.3  &  4.21  &  2.34  &  1663.8  \\
0.50   & 3.42 x 10$^{-4}$   &  1.52  &  1.52  &   587.3  &  4.27  &  2.28  &  1570.3  \\
0.60   & 4.07 x 10$^{-4}$   &  1.52  &  1.52  &   587.3  &  6.24  &  1.59  &      12.2  \\
0.70   & 4.36 x 10$^{-4}$   &  1.52  &  1.52  &   587.3  &  6.33  &  1.59  &      11.5  \\
0.80   & 4.11 x 10$^{-4}$   &  1.54  &  1.52  &   587.3  &  6.59  &  1.71  &      10.2  \\
0.90   & 3.74 x 10$^{-4}$   &  1.54  &  1.52  &   587.3  &  6.42  &  1.71  &        9.7  \\
\hline
\end{tabular}
\label{tab:statsDD}
\end{table*}

Each binary system was evolved from the ZAMS to an age of $9.5\,$Gyr,
taken to be the age of the Galactic disc \citep[e.g.][]{Oswalt1996,
  liu2000}, with the rapid binary star evolution ({\sc bse}) algorithm
developed by \citet{Hurley2002}.  This is an extension of their single
star evolution algorithm \citep{Hurley2000} in which they use
analytical formulae to approximate the full numerical hydrodynamic and
nuclear evolution of stars.  This includes mass-loss episodes during
various stages of evolution.  The {\sc bse} code adds interactions
between stars, such as mass transfer, RLOF, common
envelope evolution, supernova kicks and angular momentum loss by
gravitational radiation and magnetic braking as well as tidal
interaction. We summarize the type of stars that play a role in the
BSE code in Table\,\ref{tab:bsetypes}.

In the {\sc bse} model we use the $\alpha$ (energy) formulism for
common envelope phases and have taken a fixed $\lambda=0.5$ as
representative of the range expected for our stars.  We take $\alpha$
to be a free parameter between $0.1$ and~$0.9$.  Efficiencies of
$\alpha > 1$ are only possible if additional energy sources are
involved in the process.  We do not consider this here.  We use the
full suite of mass-loss rates described by \citet{Hurley2000}. We
found that, in order to generate sufficient low-mass white dwarfs,
$\eta=1.0$ for Reimers' mass-loss parameter is necessary so we have
used this throughout.  Alternatively sufficient low-mass white dwarfs
could be formed with smaller $\eta$ if the Galactic disc were somewhat
older. \citet{Meng2008} produce them with $\eta=0.25$ in populations
of 12\,Gyr in age. The metallicity is taken to be solar ($Z = 0.02$)
in all our calculations.

From all evolved systems we select those that could generate single
HFMWDs. To this end we select all pairs of white dwarfs that merge
outside any common envelope and leave a single white dwarf
remnant. These are our white dwarf-white dwarf (DD) mergers. Added to
these are white dwarf remnants of systems that underwent at least one
common envelope phase and merged during the last common envelope phase
and satisfy two further criteria. Firstly, either one or both of the
stars must have a degenerate core before merging and secondly, there
must be no further core burning before the remnant white dwarf is
exposed. We assume that such a core burning would be convective and
destroy any frozen-in high magnetic field.

\section{Population synthesis results}

\begin{figure*}
\begin{center}
\includegraphics[bb=-10 1 710 750,width=1.2\textwidth]{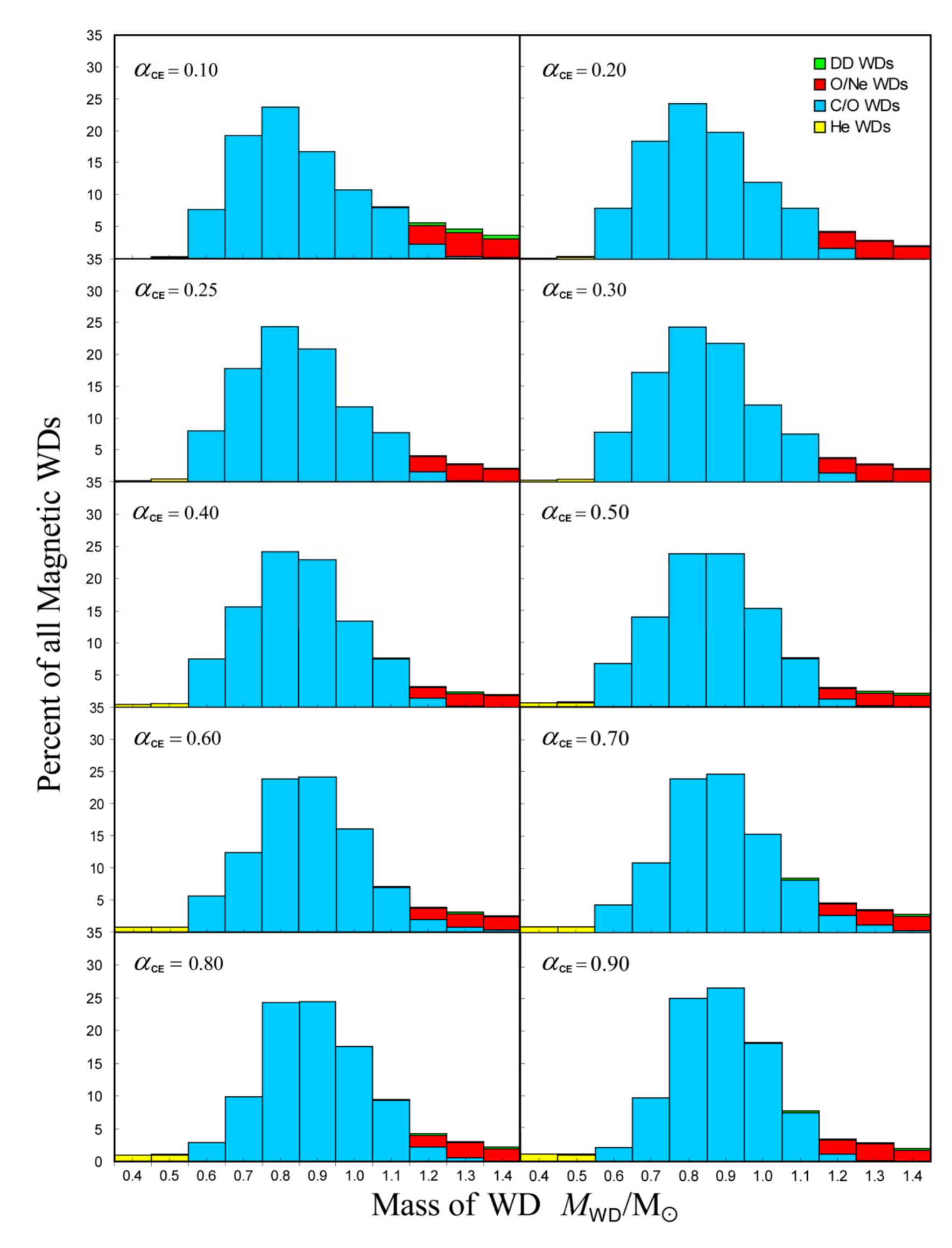}
\caption{Theoretical mass distribution of remnant white dwarfs formed
  by merging for a range of values $\alpha$ and a Galactic disc age of
  9.5\,Gyr. `DD white dwarfs' are white dwarfs resulting from DD
  mergers, `ONe white dwarfs' are Oxygen-Neon white dwarfs, `CO white
  dwarfs' are Carbon-Oxygen white dwarfs and `He white dwarfs' are
  helium white dwarf remnants after merging. This figure will appear
  in colour in the on-line version of this paper making it clearer as
  to what are the components of the high mass tail of the
  distribution.}
\label{fig:WDmasses}
\end{center}
\end{figure*} 

Assuming a constant star formation rate, each synthetic population was
integrated to the Galactic disc age over the entire parameter space
with $0.05\le\alpha\le0.9$.  Table~\ref{tab:statsCE} lists the
fraction by type of all binary systems that merge in a common envelope
and Table~\ref{tab:statsDD} those that merge as DDs in a single
generation of stars of age 9.5\,Gyr. The tables also show the limits
of the parameter space within which the cores merge.  The minimum ZAMS
masses of the systems that merged and ended their lives as single
white dwarfs are listed in the columns $M_{1_{\rm min}}$ and
$M_{2_{\rm min}}$ and the minimum initial period in the column
$P_{0_{\rm min}}$.  The maximum ZAMS masses and period are shown in
the columns $M_{1_{\rm max}}$, $M_{2_{\rm max}}$ and $P_{0_{\rm
    max}}$.  For systems that merge during a common envelope phase the
minimum ZAMS primary mass is determined by the age of the Galactic
disc and thus by the time taken by this star to evolve off the main
sequence.  For the DD route both stars must evolve to white dwarfs.
After the last common envelope episode, the two stars continue their
evolution to the white dwarf final stage.  The stars are then brought
together by gravitational radiation and eventually coalesce.  This
process takes longer than the common envelope route.  As a
consequence, the main-sequence evolution lifetime of the primary star
must be shorter and thus the minimum ZAMS mass must be larger than
that required to merge during common envelope evolution.  Otherwise
such systems would not be able to coalesce within the age of the
Galactic disc.

\begin{table}
\caption{Types and fractions per cent of white dwarfs
    formed from common envelope and DD by merging binary systems  in a population aged 9.5\,Gyr.  All DD white dwarfs are of CO type.}
\centering
\begin{tabular} { c r r r r c }
\hline
$\alpha$  &  \multicolumn{3}{c}{Common}  &  {Double}   \\
&   \multicolumn{3}{c}{envelope}  &   {degenerate} \\
& He &   CO  & ONe & CO  \\
\hline
0.05  &  0.04 &  88.77 & 11.04 &	0.15  \\
0.10  &  0.16 & 88.73  &   9.79 & 1.32  \\
0.20  &  0.43 & 92.14  &   7.08 & 0.36  \\
0.25  &  0.55 & 92.24  &   6.74 & 0.47  \\
0.30  &  0.68 & 92.10  &   6.63 & 0.58  \\
0.40  &  0.94 & 92.89  &   5.42 & 0.75  \\
0.50  &  1.20 & 92.55  &   5.41 & 0.84  \\
0.60  &  1.45 & 91.70  &   5.95 & 0.89  \\
0.70  &  1.68 & 91.20  &   6.27 & 0.85  \\
0.80  &  1.92 & 91.12  &   6.12 & 0.84  \\
0.90  &  2.20 & 90.42  &   6.47 & 0.91  \\
\hline
\end{tabular}
\label{tab:WDtypes}
\end{table}

For low values of $\alpha$ the envelope clearance efficiency is low and the time
for the envelope to exert a drag force on the orbit is largest.
Correspondingly, Table\,\ref{tab:statsCE} shows that, for low
$\alpha$, the number of coalescing stars in the common envelope path
is maximal.  As $\alpha$ increases, the time for ejection of the
envelope decreases and the number of systems that merge while still in
the common envelope also decreases.  White dwarfs formed from merged
stars are of the three types He, CO and ONe.  The small fraction of He
white dwarfs increases with $\alpha$ while that of the ONe white
dwarfs falls.  The He white dwarfs originate when RGB stars coalesce
with very low-mass main-sequence stars.  At low $\alpha$ these stars
merge when there is very little envelope left and the resulting giant
can lose the rest of its envelope before helium ignition.  As $\alpha$
is increased, more of the envelope remains after coalescence and the
stars pass through core helium burning before being exposed as CO
white dwarfs.  The ONe white dwarfs form when the most evolved AGB
stars coalesce with their companions.  These stars have only rather
weakly bound envelopes so that as $\alpha$ is increased more of them
emerge from the common envelope phase detached.  For the DD case we
find that only CO white dwarfs are formed in our models.
Table~\ref{tab:WDtypes} sets out the types and fractions of all white
dwarfs that form from common envelope and DD merging systems as a
function of $\alpha$.  The lack of merged He white dwarfs seems to
indicate that, while it is true that very low-mass white dwarfs ($M
\lsimeq 0.4\,\rm{\msun})$ must arise from binary interaction, they do
not arise from DD mergers within a Galactic disc age of 9.5\,Gyr.

\subsection{Example evolutionary histories}

The precise evolutionary history of a binary system depends on
its particular parameters.  For example the number of common envelope
events that can occur can vary from one to several \citep{Hurley2002}.  Here we give a few examples to
illustrate the difference between common envelope and DD merging events.

\subsubsection{Common envelope coalescence}

\begin{table*}
  \caption{Evolutionary history of an example binary system that
    merges during common envelope.  Here $\alpha=0.2$, $P_0= 219.6\,$d,
    S1 is the primary star and S2 is the secondary star.}  \centering
\begin{tabular}{ c r c c c l }
\hline
Stage  & Time/Myr & $M_1/\msun$ & $M_2/\msun$ & $a/{\rm R}_\odot$ & Remarks \\
\hline
1    &      0.0000    &  4.444   &   0.719   &   264.679   &   ZAMS \\
2    &  138.1295    &  4.444   &   0.719   &   264.679   &   S1 becomes a Hertzsprung gap star \\
3    &  138.7479    &  4.444   &   0.719   &   264.739   &   S1 becomes a red giant  \\
4    &  139.1676    &  4.443   &   0.719   &   179.877   &   S1 starts core helium burning. Some mass loss occurs \\
5    &  161.7637    &  4.402   &   0.719   &   181.495   &   S1 first AGB \\
6    &  161.9691    &  4.402   &   0.719   &   141.380   &   S1 begins RLOF \\
7    &  161.9691    &  4.524   &           -   &  { } 0.529   &   common envelope: S1, S2 coalesce; RLOF ends \\  
8    &  162.8725    &  4.494   &           -   &               -   &   S1 becomes late AGB \\
9    &  163.5543    &  0.924   &           -   &               -   &   S1 becomes CO white dwarf \\
\hline
\end{tabular}
\label{tab:evolCE}
\end{table*}

Table~\ref{tab:evolCE} sets out the evolutionary history of an
example system that merges during a common envelope with $\alpha=0.2$.
The progenitors are a primary star S1 of $4.44\rm\,\msun$ and a
secondary S2 of sub-solar mass $0.72\rm\,\msun$.  At ZAMS the initial
period is $219.6\,$d and the orbit is circular with a separation of
$264.7\,\rm R _\odot$.  S1 evolves first and reaches the early AGB at 161.77\,Myr
having lost $0.02\,\rm\msun$ on the way.  Roche lobe overflow starts 0.2\,Myr
later with mass flowing from S1 to S2.  At this point the orbital separation
has decreased to 141.4$\,\textrm{R}_\odot$ because orbital angular momentum
has been lost through tidal spin up of S1.  A common envelope develops and the two cores
coalesce when their separation reaches $0.53\,\textrm{R}_\odot$.
A further $0.6\,\rm\msun$ of the envelope has been lost.  At 162.78\,Myr,
approximately 0.9\,Myr after coalescing, S1 becomes a late stage AGB star.
After a further 0.7\,Myr it becomes a CO white dwarf.

\subsubsection{DD coalescence}

\begin{table*}
  \caption{Evolutionary history of an example of white dwarf that formed
    in a double degenerate coalescence.  Here $\alpha=0.1$,
    $P_0= 3144$\,days, S1 is the primary star and S2 the secondary
    star.}
    \centering
\begin{tabular}{ c r c c r l }
\hline
Stage  & Time/Myr & $M_1/\msun$ & $M_2/\msun$ & $a/{\rm R}_\odot$ & Remarks \\
\hline
1   &         0.0000   &   3.673   &   1.928   &   1603.362   &   ZAMS \\
2   &     222.4734   &   3.673   &   1.928   &   1603.362   &   S1 becomes a Hertzsprung gap star \\
3   &     223.6164   &   3.673   &   1.928   &   1603.416   &   S1 becomes a Red Giant \\
4   &     224.6021   &   3.672   &   1.928   &   1603.678   &   S1 starts core helium burning \\
5   &     268.5530   &   3.645   &   1.928   &   1611.505   &   S1 becomes early AGB \\
6   &     270.4541   &   3.614   &   1.928   &   1583.219   &   S1 becomes late AGB \\
7   &     270.9681   &   2.682   &   1.947   &   1509.115   &   S1 begins RLOF, mass transfer on to S2 and mass loss occurs \\
8   &     270.9681   &   0.821   &   1.947   &     374.233   &   common envelope. S1 emerges as a CO white dwarf and RLOF ends \\
9   &   1260.0681   &   0.821   &   1.947   &     374.233   &   Begin Blue Straggler phase \\
10 &   1267.0548   &   0.821   &   1.947   &     374.233   &   S2 becomes a Hertzsprung gap star \\
11 &   1277.4509   &   0.821   &   1.946   &     374.245   &   S2 becomes a Red Giant \\
12 &   1306.9423   &   0.821   &   1.943   &     375.353   &   S2 starts core helium burning \\
13  &  1513.3615   &   0.821   &   1.926   &     377.768   &   S2 becomes early AGB \\
14  &  1517.2953   &   0.821   &   1.913   &     324.600   &   S2 begins Roche lobe overflow \\
15  &  1517.2953   &   0.821   &   0.536   &         2.433   &   common envelope begins. S2 evolves to a CO white dwarf and RLOF ends \\
16  &  9120.8467   &   0.821   &   0.536   &         0.040   &   S2 begins RLOF\\
17  &  9120.8467   &   1.357   &          -    &         0.000   &   S1, S2 coalesce \\
\hline
\end{tabular}
\label{tab:evolDD}
\end{table*}

In the DD pathway both stars survive the common envelope without
merging and both continue to evolve to white dwarfs approaching each
other through gravitational radiation to eventually coalesce.
Table~\ref{tab:evolDD} illustrates this for $\alpha=0.1$.  At ZAMS the
progenitors are a $3.7\,\rm\msun$ primary and a $1.9\,\rm\msun$
secondary with an initial period of 3\,444\,d and a separation of
$1603\,\textrm{R}_\odot$, again in a circular orbit.  The primary
evolves through to a late stage AGB star after 270.5\,Myr losing
$0.6\,\rm\msun$ on the way. The separation falls
to~$1509\,\textrm{R}_\odot$.  As a late AGB star S1 loses
$0.9\,\rm\msun$ of which $0.02\,\rm\msun$ is accreted by S2 from the
wind.  Approximately 0.5\,Myr later, at 271\,Myr with S1\, of mass
$2.68\,\rm\msun$ and S2 $1.95\,\rm\msun$, RLOF commences and a common
envelope develops.  The orbital separation falls to
$374\,\rm\textrm{R}_\odot$ when the envelope is ejected.  S2 continues
to evolve, first as a blue straggler then through the Hertzsprung gap,
red giant and core helium burning stages until it becomes an early AGB
star at 1513.4\,Myr.  At 1517.3\,Myr RLOF begins again and a second
common envelope forms.  At an orbital separation of only
$2.43\,\rm{R}_\odot$ the envelope is ejected and S2 emerges as a CO
white dwarf of mass $0.54\,\rm\msun$.  A long period of orbital
contraction by gravitational radiation follows until at 9120.8\,Myr
the two white dwarfs are separated by $0.04\,\textrm{R}_\odot$ and
RLOF from S2 to S1 begins followed rapidly by coalescence of the DDs.
The remnant star is still a CO white dwarf but now of mass
$1.36\,\rm\msun$.

\subsection{Mass distribution of the synthetic population}

\begin{figure}
\begin{minipage}{83mm}
\begin{center}
\includegraphics[bb=5 10 900 256, width=2.3\textwidth]{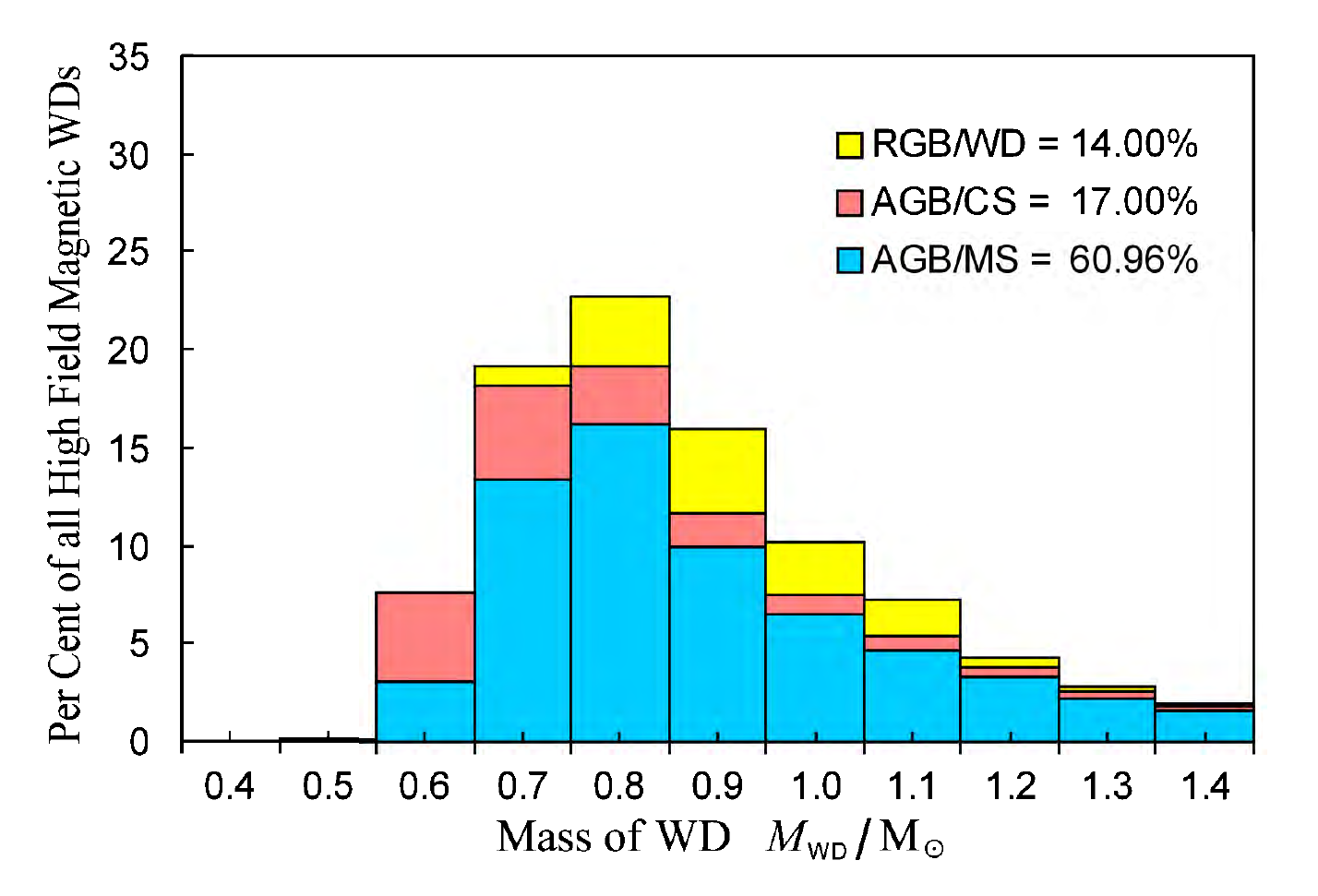}
\caption{Mass distribution of theoretical HFMWDs for $\alpha=0.10$ separated
  according to their pre-common envelope progenitors.  Other paths also contribute but
  are less than 1\,per cent of the total.  The Galactic disc age is
  chosen to be 9.5\,Gyr.  The stellar types are identified in
  Table~\ref{tab:bsetypes}. This figure will appear in colour in the on-line version
    of this paper making it clearer as to what are  the various paths.}
\label{fig:Path}
\end{center}
\end{minipage}
\end{figure} 

With the selected common envelope and DD merged systems we generate a
population of putative magnetic white dwarfs by integration over time
from $t=0$ to $9.5\,$Gyr, our chosen age for the Galactic disc.  The
star formation rate is taken to be constant over the lifetime of the
Galactic disc.  Whereas Tables\,\ref{tab:statsCE} and
\ref{tab:statsDD} show the relative numbers of merged white dwarfs
from a single generation of binary stars, continuous star formation
over the lifetime of the Galaxy builds up a population of white dwarfs
that favours higher-mass systems because at lower-mass, especially in
later generations, they do not have enough time to evolve.  Similarly,
the slow orbital contraction by gravitational radiation means that
potential DD coalescence in later generations is not complete and the
fraction of those white dwarfs is further reduced in the present day
population.  Fig.~{\ref{fig:WDmasses}} shows the mass distribution for
CO, ONe and DD white dwarfs in a present day population formed over
the age of the Galactic disc, 9.5\,Gyr.  Fig.~{\ref{fig:Path}} shows
the contributions from the various pre-common envelope progenitor
pairs that formed the post-common envelope white dwarfs either through
the common envelope or DD path when $\alpha$ = 0.1.  Other paths also
contribute but to less than 3\,per cent of the total each.
Table~\ref{tab:Contribs} lists their contributions summed over all
white dwarf masses.

\begin{table}
\caption{The contributions per cent of pre-common envelope progenitor pairs to theoretical
 HFMWDs when $\alpha = 0.1$. The stellar type `CS' is a deeply
 or fully convective low-mass main sequence star (see Table \ref{tab:bsetypes}).}
\centering
 \label{tab:Contribs}
\begin{tabular}{c c}
\hline
Progenitor pairs  & Fraction per cent\\
\hline
AGB/MS	           & 60.96  \\
AGB/CS	           & 17.00  \\
RGB/CO WD	           & 14.00  \\
AGB/HG                   & {  }2.72  \\
AGB/CO WD            & {  }2.21  \\
CO WD/CO WD       & {  }1.32  \\
RGB/RGB	           & {  }0.97  \\ 
RGB/AGB	           & {  }0.46  \\
RGB/CS                    & {  }0.16  \\
AGB/AGB                  & {  }0.20  \\
\hline
\end{tabular}
\end{table}

In order to calculate the incidence of HFMWDs we used the same
{\sc{bse}} code to model single star evolution through to the white dwarf stage
also for a Galactic disc age of 9.5\,Gyr under the assumption that all white dwarfs originating
from single star evolution are non-magnetic.. Table\,\ref{tab:MagWD_perc} sets
out the incidence of HFMWDs as a percentage of the incidence of field
white dwarfs for a range of $\alpha$.

\section{Comparison with observations}
\label{sec:Compare}

\begin{table}
\caption{The theoretical incidence of  HFMWDs as a fraction of 
magnetic to non-magnetic field white dwarfs as a function of
  the common envelope efficiency parameter $\alpha$.}
\label{tab:MagWD_perc}
\centering
\begin{tabular} { l r c r}
\hline
$\alpha$ & \multicolumn{3}{c}{HFMWDs per cent} \\
&   {CE} & {DD} & {Total}  \\
\hline
0.05   & 21.63 & 3.16 x 10$^{-2}$  &   21.67  \\
0.10	& 18.99 & 2.58 x 10$^{-1}$  &  19.25  \\
0.20	& 16.12 & 5.80 x 10$^{-1}$  &  16.18  \\
0.25   & 14.78 & 7.02 x 10$^{-2}$  &  14.85  \\
0.30	& 13.50 & 8.03 x 10$^{-2}$  &  13.58  \\
0.40	& 11.85 & 8.80 x 10$^{-2}$  &  11.67  \\
0.50	& 10.10 & 8.64 x 10$^{-2}$ &  10.18  \\
0.60   &   8.94 &  8.11 x 10$^{-2}$ &    9.02  \\
0.70	&   8.15 &  7.01 x 10$^{-2}$  &    8.22  \\
0.80   & 18.99 &  6.33 x 10$^{-2}$  &    7.50  \\
0.90	& 18.99 &  6.24 x 10$^{-2}$  &    6.78  \\
\hline
\end{tabular}
\end{table}

We compare our theoretical predictions with observations of HFMWDs.
Our comparison includes (i)~the incidence of magnetism among single
white dwarfs and (ii)~the mass distribution of single HFMWDs.  This is
not a simple task because the observational data base of HFMWDs is a
mixed bag of objects from many different ground and space-borne
surveys.  It is plagued by observational biases.  In magnitude-limited
surveys, such as the Palomar-Green (PG) or the Hamburg-Schmidt
surveys, one of the biases against the detection of magnetic white
dwarfs has been that since these are generally more massive than their
non-magnetic counterparts \citep{Liebert1988}, their radii are smaller
and therefore they are less luminous.  Similar biases would also
apply to UV and X-ray surveys.  However \citet{Liebert2003} have
argued that, in any \emph{explicitly magnitude-limited survey}, it may
be possible to correct for the difference in search volume for the
magnetic white dwarfs.  Thus a re-analysis of the data of the PG
survey, that took into account the different volumes that are sampled
by different mass white dwarfs, gave an estimate for the fraction of
HFMWDs of at least $7.9\pm 3$\,per cent \citep{Liebert2003}.
Volume-limited samples are expected to be less affected by the radius
bias but contain very few magnetic white dwarfs with known masses or
temperatures.  A nearly complete volume-limited sample of nearby white
dwarfs by \citet{Kawka2007} shows that up to 21$\,\pm\,$8\,per cent of
all white dwarfs within 13\,pc have magnetic fields greater than about
3\,kG and 11$\,\pm\,$5\,per cent are HFMWDs with $B\ge 1$\,MG. 

\begin{figure}
\begin{minipage}{83mm}
\begin{center}
\includegraphics[bb= 5 10 480 240,width=1.2\textwidth]{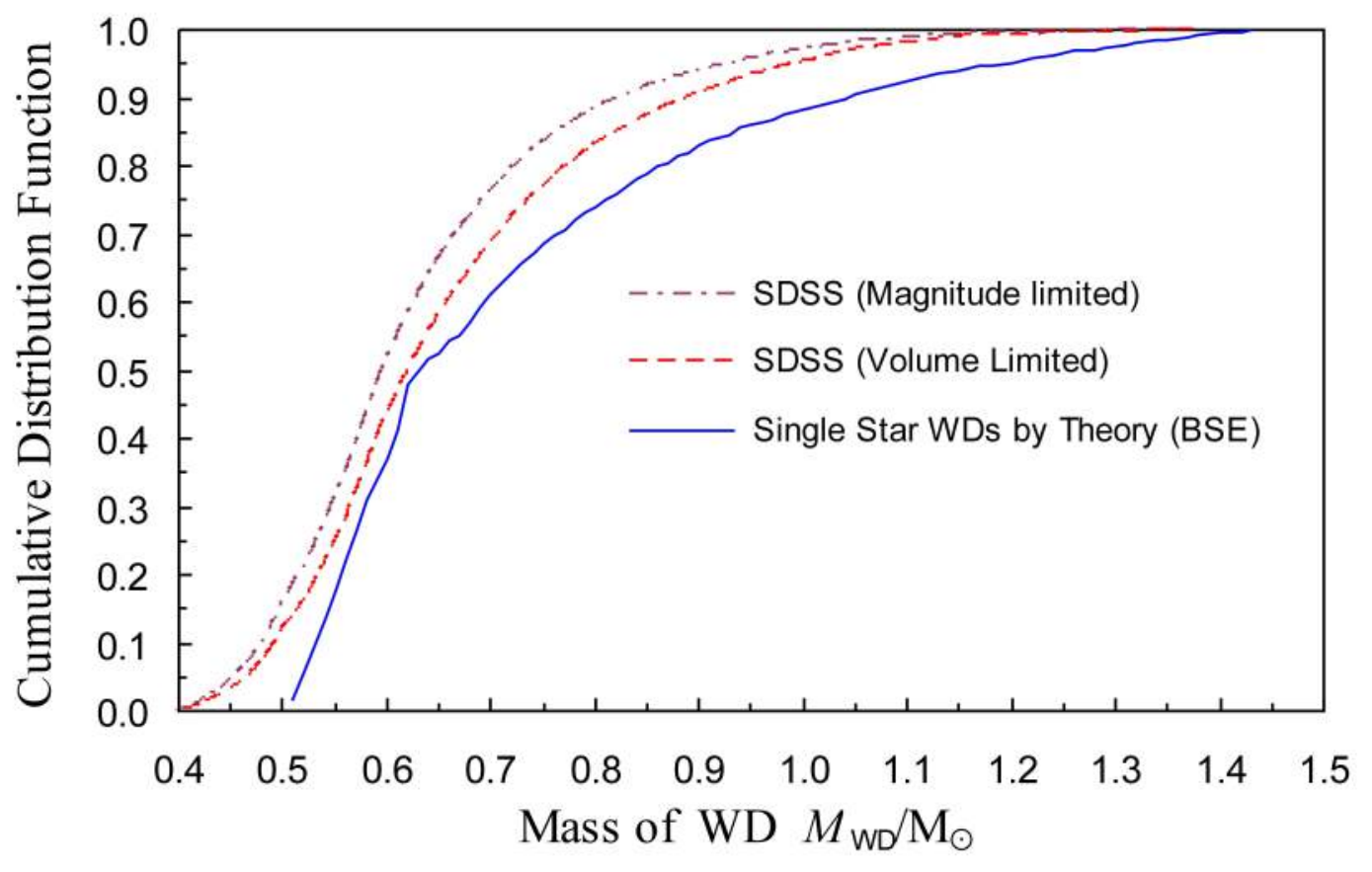}
\caption{CDFs of masses of observed SDSS
  DR7 \citep{Kleinman2013} non-magnetic, magnitude-limited and
  converted-volume-limited field white dwarfs and the theoretical
  ({\sc{bse}}) volume-limited population of non-magnetic white dwarfs
  from single star evolution for a Galactic disc age of 9.5\,Gyr.}
\label{fig:SingleCDF}
\end{center}
\end{minipage}
\end{figure}

The synthetic population generated by {\sc{bse}} is a volume-limited
sample and so is not directly comparable with a magnitude limited
sample such as the Sloan Digital Sky Survey Data Release\,7 (SDSS DR7)
white dwarf catalogue \citep{Kleinman2013} which has 12\,803 members.
\citet{Liebert2003} estimated that the limiting distance to which a
white dwarf can be found in a magnitude-limited survey is proportional
to its radius $R_{\rm{WD}}$.  Thus the survey volume for a given mass
scales as $R^3_{\rm{WD}}$.  We correct this bias by weighting each
white dwarf found by the SDSS in proportion to $1.0/R^3_{\rm{WD}}$
relative to the radius of a 0.8\,\msun white dwarf.  The cumulative
distribution function (CDF) for the corrected mass distribution along
with the CDF for the uncorrected mass distribution of the SDSS white
dwarfs is shown in Fig.~\ref{fig:SingleCDF}.  The theoretical CDF
obtained with {\sc bse} for the mass distribution of single white
dwarfs is shown for comparison.

We note that the {\sc{bse}} code we use does not produce low-mass
white dwarfs because of the limited age of our Galactic disc. However
\citet{Han1994} and \citet{Meng2008} have constructed single star
models using different assumptions utilizing a superwind that produces
low-mass white dwarfs in older populations. This is also reflected in
the inability of our {\sc{bse}}  results to demonstrate the
existence of a significant fraction of low-mass He white dwarfs.

\begin{table*}
\caption{Known HFMWDs with poloidal field strength $B_{\rm pol} \ge
  10^5$\,G.  In the comparison with our models we exclude five of
  these white dwarfs with $B_{\rm pol} < 1\,$MG (nos~1, 3, 18, 20 and 32) and
  two more of extremely low mass (nos~19 and 29) that cannot be
  formed within the {\sc bse} formulism.}
\label{tab:Obsdb}
\centering
\begin{tabular} { r l l c l l l }
\hline
No. & white dwarf & Aliases & $B_{\rm pol}/$MG & $T_{\rm eff}/$K & Mass/\msun &
References \\
\hline
1	& 0009+501      & LHS 1038, G217-037, GR381& $\lesssim  0.2$	& \phantom{0}6540 $\pm$ 150	& 0.74 $\pm$ 0.04		&  1,23		\\
2	& 0011--134     & LHS 1044, G158-45		& $16.7 \pm 0.6$	& \phantom{0}3010 $\pm$ 120	& 0.71 $\pm$ 0.07		&  2, 3		\\
3	& 0257+080      & LHS 5064, GR 476		& $\approx 0.3$	& \phantom{0}6680 $\pm$ 150	& 0.57 $\pm$ 0.09		&  2			\\
4	& 0325--857     & EUVE J0317-855		& $185-450$		& 33000			           & 1.34 $\pm$ 0.03		&  4			\\
5	& 0503--174     & LHS 1734, LP 777-001		& $7.3 \pm 0.2$	& \phantom{0}5300 $\pm$ 120	& 0.37 $\pm$ 0.07		&  2, 3		\\
6	& 0584--001     & G99-37			           & $\approx 10$	& \phantom{0}6070 $\pm$ 100	& 0.69 $\pm$ 0.02		&  5,6,7		\\
7	& 0553+053     & G99-47			           & $20 \pm 3$	& \phantom{0}5790 $\pm$ 110	& 0.71 $\pm$ 0.03		&  2, 7, 8		\\ 
8	& 0637+477     & GD 77				& $1.2 \pm 0.2$	& 14870 $\pm$ 120	                     & 0.69				&  9, 10		\\
9	&0745+304      & SDSS J074853.07+302543.5	& 11.4		& 21000 $\pm$ 2000	           & 0.81 $\pm$ 0.09		& 44   		           \\
10	& 0821--252     & EUVE J0823-254		& $2.8-3.5$		&  \verb++43200 $\pm$ 1000	& 1.20 $\pm$ 0.04		& 11			\\
11	& 0837+199    & EG 061$^b$, LB 393		& $\approx 3$	& 17100 $\pm$ 350	                     & 0.817$\pm$ 0.032	 & 12		           \\
12	& 0912+536    & G195-19			           & 100		           & \phantom{0}7160  $\pm$ 190	& 0.75 $\pm$ 0.02		&  2, 13, 14		\\
13	&               & SDSS J092646.88+132134.5	& $210 \pm 25$	& \phantom{0}9500  $\pm$ 500	& 0.62 $\pm$ 0.10		& 15			\\
14	& 0945+246      & LB11146$^a$			& 670			& 16000 $\pm$ 2000	           & 0.90 (+0.10, -0.14)	& 16, 17		\\
15	& 1026+117	& LHS 2273			           & 18			& \phantom{0}7160 $\pm$ 190	& 0.59			& 18		           \\
16	& 1220+234	& PG1220+234			& 3			& 26540			           & 0.81			& 19			\\
17	& 1300+590	& SDSS J13033.48+590407.0	& $\approx 6$	& \phantom{0}6300 $\pm$ 300	& 0.54 $\pm$ 0.06		& 20			\\
18	& 1328+307 	& G165-7		  	          & 0.65		           & \phantom{0}6440 $\pm$ 210	& 0.57 $\pm$ 0.17		& 21			\\
19	& 1300+015	& G62-46 			           & $7.36 \pm 0.11$	& \phantom{0}6040		& 0.25			& 22			\\
20	& 1350--090	& LP 907-037			& $\lesssim 0.3$	& \phantom{0}9520 $\pm$ 140	& 0.83 $\pm$ 0.03		& 23, 24		\\
21	& 1440+753	& EUVE J1439+750$^a$ 		& $14-16$		& 20000-50000			& 1.04 (+0.88, -1.19)	& 25			\\
22	& 1503-070	& GD 175$^a$			& 2.3			& \phantom{0}6990		& 0.70 $\pm$ 0.13	           & 2			\\
23	&               & SDSS J150746.80+520958.0	& $65.2 \pm 0.3$	& 18000 $\pm$ 1000	& 0.99 $\pm$ 0.05		           & 15			\\			
24	&               & SDSS J150813.24+394504.0	& 18.9		& 18000 $\pm$ 2000	& 0.88 $\pm$ 0.06		           & 44			\\
25	& 1533--057	& PG 1355-057			& $31 \pm 3$	& 20000 $\pm$ 1040	& 0.94 $\pm$ 0.18		           & 26, 27, 25		\\
26	& 1639+537	& GD 356, GR 329			& 13			& 7510 $\pm$ 210	& 0.67 $\pm$ 0.07		                      & 2, 28, 29, 45\\
27	& 1658+440	& 1658+440, FBS 376		& $2.3 \pm 0.2$	& 30510 $\pm$ 200	& 1.31 $\pm$ 0.02		                      & 11, 30		\\
28	& 1748+708	& G240-72			           & $\gtrsim 100$	& \phantom{0}5590 $\pm$ 90	& 0.81 $\pm$ 0.01		& 2, 5			\\
29	& 1818+126	& G141-2$^a$			& $\approx 3$	& \phantom{0}6340 $\pm$ 130	& 0.26 $\pm$ 0.12		& 18, 31		\\
30	& 1829+547	& G227-35			           & $170-180$		& \phantom{0}6280 $\pm$140	& 0.90 $\pm$ 0.07		& 2, 8			\\
31	& 1900+705	& AC +70$^\circ$8247, GW +70$^\circ$8247 & $320 \pm 20$	& 16000		& 0.95 $\pm$ 0.02		& 2, 32, 33, 34, 35, 36	\\
	& 		& EG 129, GL 742, LHS 3424\\	
32	& 1953--011	& G92-40, LTT 7879, GL 772	& $0.1-0.5$		& \phantom{0}7920 $\pm$ 200	& 0.74 $\pm$ 0.03		& 2, 37, 38		\\
	&		& LP 634-001, EG 135, LHS 3501\\
33	& 2010+310	& GD 229, GR 333		           & $300-700$		& 16000			           &  1.10-1.20			& 33, 35, 39, 40, 41, 42\\
34	& 2329+267	& PG 2329+267, EG 161		& $2.31 \pm
0.59$	& \phantom{0}9400 $\pm$ 240	&  0.61 $\pm$ 0.16
& 2, 43, 24		\\
\\

\hline
\end{tabular}

\begin{tabular} {l}
\begin{minipage}{168mm}
$^a$ Unresolved DD\\
$^b$ Praesepe (M44, NGC 2632)\\ \\
\verb+   +\textbf{References:}
  (1) \citet{Valyavin2005}; 
  (2) \citet{Bergeron2001}; 
  (3) \citet{Bergeron1992}; 
  (4) \citet{Vennes 2003}; 
  (5) \citet{Angel1978}; 
  (6) \citet{Dufour2005}; 
  (7) \citet{Pragal1989}; 
  (8) \citet{Putney1995}; 
  (9) \citet{Schmidt92}; 
(10) \citet{Giovannini1998}; 
(11) \citet{Ferrario1998}; 
(12) \citet{Vanlandingham2005}; 
(13) \citet{Angel1977}; 
(14) \citet{Angel1972}; 
(15) \citet{Dobbie2012}; 
(16) \citet{Glenn1994}; 
(17) \citet{Liebert1993}; 
(18) \citet{Bergeron1997}; 
(19) \citet{Liebert2003}; 
(20) \citet{Girven2010}; 
(21) \citet{Dufour2006}; 
(22) \citet{Bergeron1993}; 
(23) \citet{SchmidtSmith1994}; 
(24) \citet{Liebert05}; 
(25) \citet{Vennes 1999}; 
(26) \citet{Liebert1985}; 
(27) \citet{Achilleos1989}; 
(28) \citet{Ferrario1997a}; 
(29) \citet{Brinkworth2004}; 
(30) \citet {Schmidtetal92}; 
(31) \citet{Greenstein1986}; 
(32) \citet{Wickramasinghe1988}; 
(33) \citet{wickramasinghe2000};  
(34) \citet{Jordan1992}; 
(35) \citet{Angel1985}; 
(36) \citet{Greenstein1985}; 
(37) \citet{Maxted2000}; 
(38) \citet{Brinkworth2005}; 
(39) \citet{Green1981}; 
(40) \citet{Schmidt1990}; 
(41) \citet{Schmidt1996}; 
(42) \citet{Jordan1998}; 
(43) \citet{Moran1998}; 
(44) \citet{Dobbie2013}. 
(45) \citet{Ferrario1997a}. 
\end{minipage}
\end{tabular}
\end{table*}

From a theoretical point of view the problem of the determination of
surface gravities and masses from line spectra of HFMWDs has also
proved to be insoluble, except for low-field objects ($B \lsimeq
3\,$MG) for which we can assume that the magnetic field does not
affect the atmospheric structure.  In these objects the field
broadening is negligible and standard zero-field Stark broadening
theories can be used to calculate the line wings
\citep[e.g.][]{Ferrario1998} and thus to determine the mass of the
magnetic white dwarf.  In principle it should also be possible to use
stationary field components that are insensitive to field structure to
estimate gravities from line profiles for HFMWDs.  Regrettably this is
not yet possible because a full theory of Stark broadening in the
presence of crossed electric and magnetic fields \citep{Main1998} has
not yet been developed.  For now, reliable mass determinations are
only available for a few low-field magnetic white dwarfs, for magnetic
white dwarfs which have good trigonometric parallaxes and magnetic
white dwarfs with white dwarf companions whose atmospheric parameters
can be established
\citep[e.g. RE\,J0317-853,][]{Barstow1995,Ferrario1997b}.  Currently
there are 34~known magnetic white dwarfs with reasonably accurately
determined masses with magnetic fields stronger than $10^5$\,G.  These
are listed in Table~\ref{tab:Obsdb} with their poloidal magnetic field
strengths, effective temperatures, masses and references in the
literature.  If we restrict ourselves to the HFMWDs with $B > 1\,$MG
we end up with 29~objects.  When comparing with our models we exclude
a further two extremely low-mass white dwarfs because it is not
possible to form these within the {\sc bse} formulism.  The most
recent additions to this list are the two common proper motion pairs
from the SDSS reported by \citet{Dobbie2013}.  We shall test our
hypothesis on this restricted mass sample with the caveat that we may
well be still neglecting observational biases. We also note that the
observational sample is neither volume nor magnitude limited.

\begin{figure}
\begin{minipage}{83mm}
\begin{center}
\includegraphics[bb=10 20 880 340,width=1.62\textwidth]{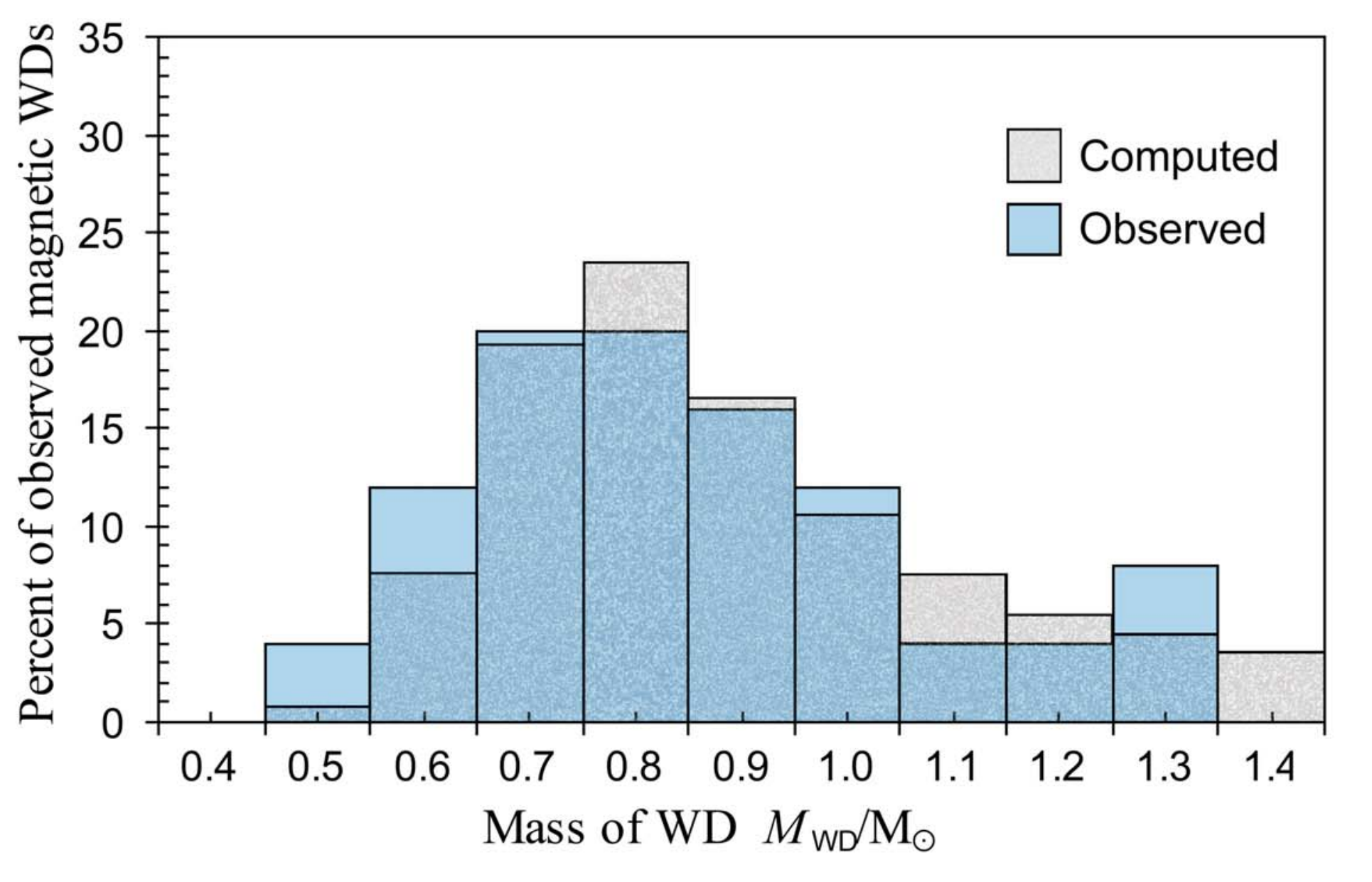}
\caption{Mass distribution of 27 observed HFMWDs (objects taken
  fromTable \ref{tab:Obsdb}) compared with the computed sample.}
\label{fig:ObsMagWDs}
\end{center}
\end{minipage}
\end{figure} 

The comparison of the mass distribution between theory and
observations is shown in Fig. \ref{fig:ObsMagWDs}.  Most of our models
reproduce the observed peak near $0.8\,\rm\msun$ but are less
successful at reproducing the higher and lower mass tails.
Interestingly the peak is dominated by giant cores that merge with
main-sequence stars.  This case was not considered by
\citet{garcia2012} who focused only on merging DDs.  We have used a
Kolmogorov-Smirnov (K--S) test \citep{Press1992} to compare the mass
distribution of the observed HFMWDs with our synthetic populations.
The K--S test determines the statistical probability that two sample
sets are drawn from the same population.  It uses the CDFs of the two
sample sets which naturally agree at the smallest value of an
independent variable where they are both zero and again at its maximum
where they are both unity.  The test then uses the intervening
behaviour to distinguish the populations.  The test gives a statistic
$D$ which is the maximum of the absolute difference between two CDFs
at a given $M_{\rm WD}$ and the probability $P$ that a random
selection from the population would lead to a larger $D$ than that
measured.

\begin{figure}
\begin{minipage}{83mm}
\begin{center}
\includegraphics[bb=45 20 400 250,width=0.87\textwidth]{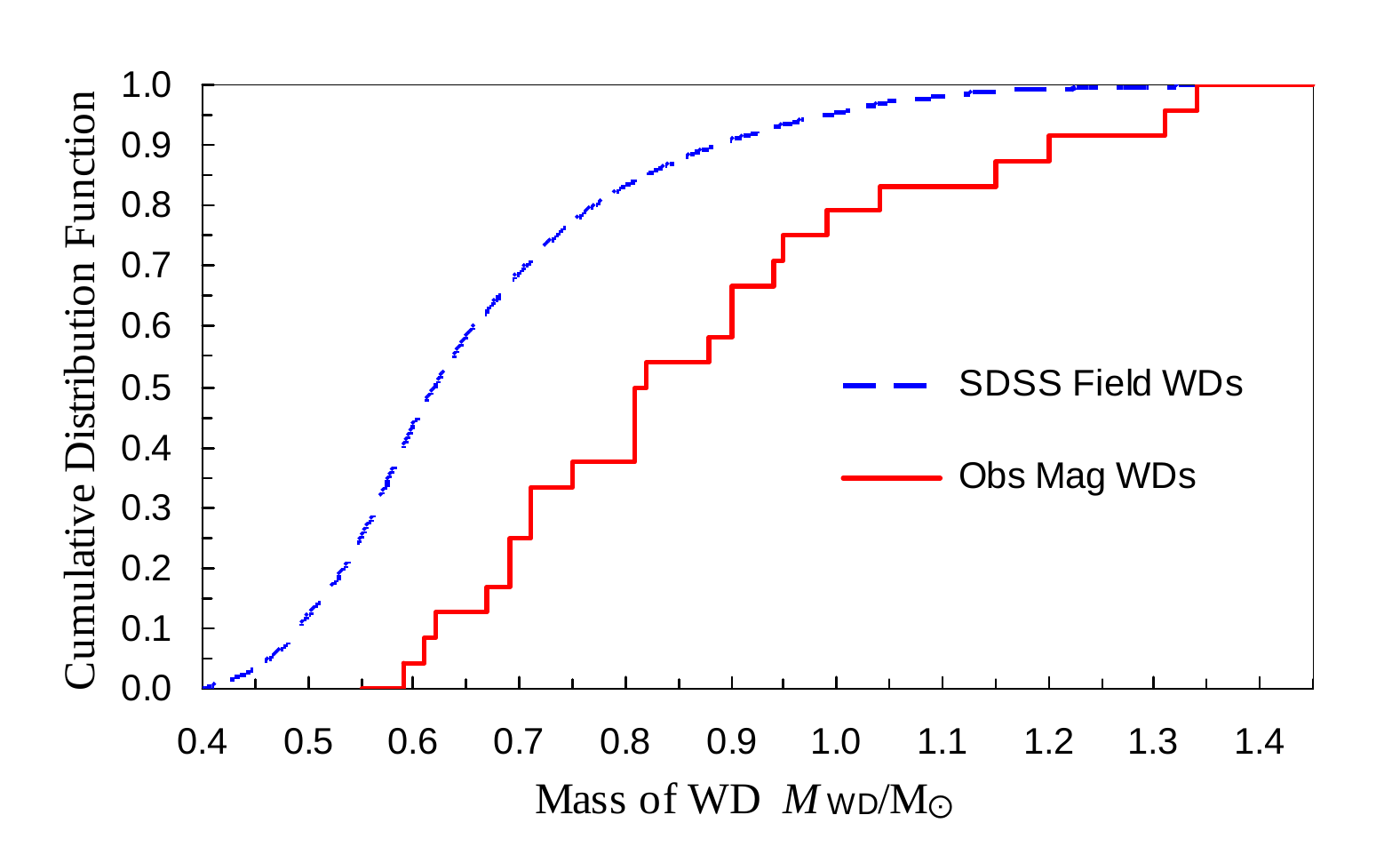}
\caption{CDFs of volume-limited-converted
  masses of observed SDSS DR7 \citep{Kleinman2013} non-magnetic, field
  white dwarfs and the observed magnetic white dwarfs.  The population of observed magnetic white dwarfs
  is not strictly a volume limited sample since it comes from various surveys as discussed in the text.
  A formal application of the K--S test has $D=0.4417$ and  $P = 3\times 10^{-5}$.}
\label{fig:MagNonMagCDF}
\end{center}
\end{minipage}
\end{figure} 

\begin{figure}
\begin{minipage}{83mm}
\begin{center}
\includegraphics[bb=45 20 400 260,width=0.87\textwidth]{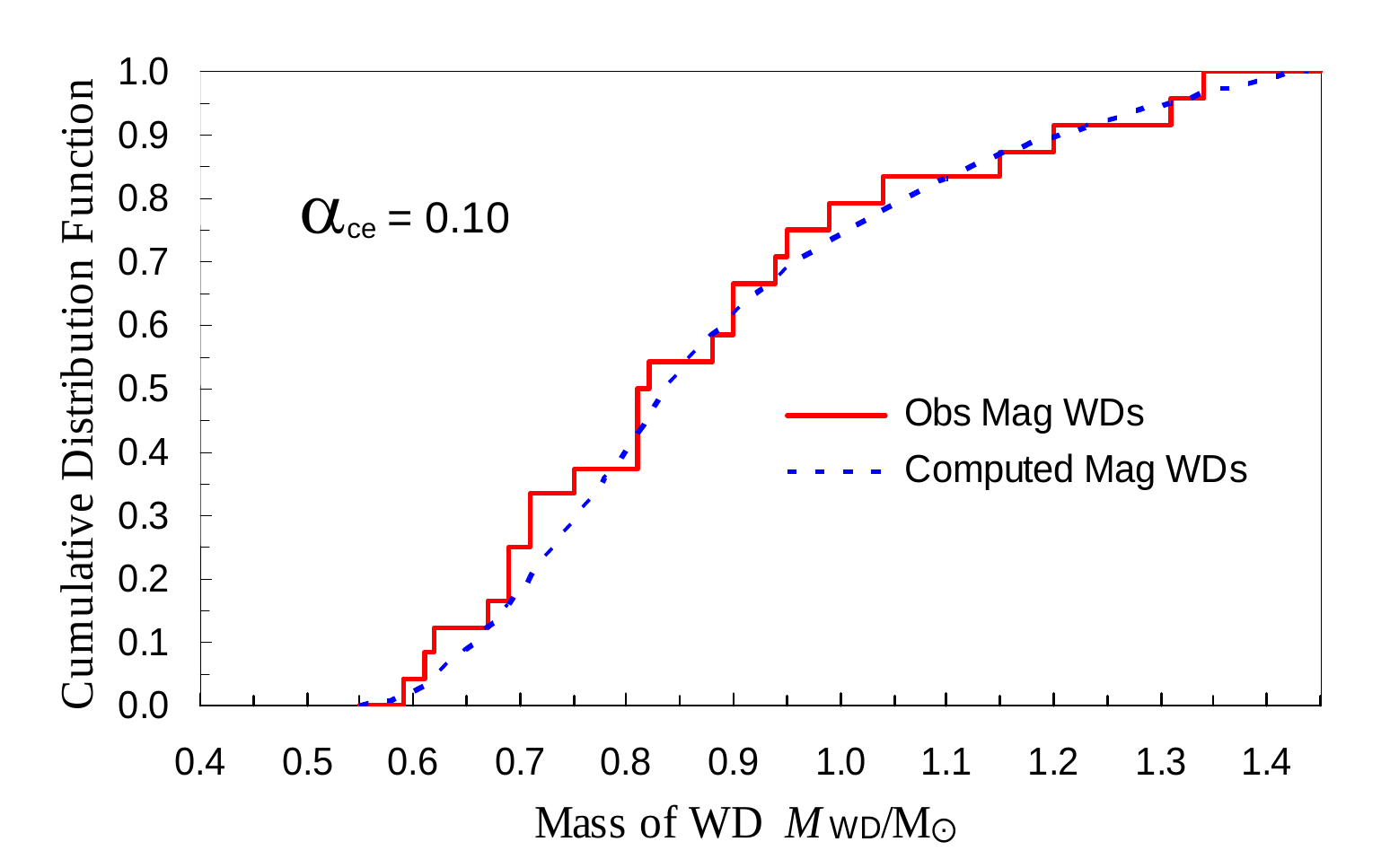}
\caption{CDF of observed and {\sc{bse}}
  theoretical HFMWD masses for a Galactic disc age of 9.5\,Gyr and
  $\alpha=0.10$.  The K--S test has $D=0.1512$ and $P=0.7095$.
  \label{fig:CDFObsTheory}}
\end{center}
\end{minipage}
\end{figure} 

\begin{table}
\caption{Kolmogorov-Smirnov $D$ statistic and $P$ of the mass
    distributions of the theoretical ({\sc{bse}}) and observed
    magnetic white dwarf populations being drawn from the same
    distribution for various values of $\alpha$. The theoretical
    population is for a Galactic disc age of 9.5\,Gyr. }
\centering

\begin{tabular} { c c c }
\hline
$\alpha$  & $D$ & $P$ \\
\hline

0.05  &  0.1558  &  0.6735  \\
0.10  &  0.1512  &  0.7095  \\
0.20  &  0.1565  &  0.6684  \\
0.25  &  0.1616  &  0.6288  \\
0.30  &  0.1675  &  0.5824  \\
0.40  &  0.1827  &  0.4700  \\
0.50  &  0.2040  &  0.3326  \\
0.60  &  0.2304  &  0.2039  \\
0.70  &  0.2580  &  0.1144  \\
0.80  &  0.2814  &  0.0665  \\
0.90  &  0.2915  &  0.0518  \\

\hline
\end{tabular}
\label{tab:KSTest}
\end{table}
Fig.~\ref{fig:MagNonMagCDF} shows the mass distribution CDFs for the
27observed HFMWDs (jagged line) and for the 12\,803 SDSS DR7 field
white dwarfs (smooth curve).  A visual inspection shows the two CDFs
to be distinctly different. The K--S test gives a $D = 0.4417$ and $P
= 3\times 10^{-5}$. So we deduce that HFMWD masses are not distributed
in the same manner as non-magnetic single white dwarfs.  When the CDF
for the observed HFMWD mass distribution is compared to the CDF for
the {\sc{bse}} theoretical mass distribution
(Fig.~\ref{fig:CDFObsTheory}) for $\alpha=0.10$ it can be seen that
the two curves are remarkably similar.  The K--S test gives a smaller
$D$ of 0.1512 with a probability of 0.7095 that indicates success of
our model.  The results of the K--S test for a range of $\alpha$s
(Table~\ref{tab:KSTest}) show that the mass distribution is consistent
over the wide range $0.05 \le \alpha\le 0.7$.  On the other hand,
based on the results in Table~\ref{tab:MagWD_perc} the observed
incidence of magnetism, as observed in the \citet{Kawka2007}
volume-limited sample, constrains $\alpha$ to be in the narrower range
$0.1 \le \alpha \le 0.3$.

\section{Discussion and Conclusions}
\label{DandC}

Two competing models for the origin of strong magnetic fields in white
dwarfs are broadly the fossil field model and the merging star model.
The proponents of the fossil field model have noted that the maximum
poloidal flux observed in the magnetic Ap and Bp stars is similar to
the maximum poloidal magnetic flux observed in the magnetic white
dwarfs.  The two groups of stars could therefore be evolutionarily
linked.  However, to date, there have been no stellar evolution models
that have shown how a strong fossil magnetic flux can survive through
the various stages of stellar evolution through to the white dwarf
phase.  It is also not clear if the similarities in the maximum
magnetic fluxes between two groups of stars is necessarily a reason to
assume a causal link.  The dynamo model of \citet{wickramasinghe2014}
suggests that similar maximum magnetic fluxes may be expected for
physical reasons if the fields are generated from differential
rotation caused by merging.  Here we have explored the consequences of
such a hypothesis for the origin of the HFMWDs with binary population
synthesis under standard assumptions, discussed in section
\ref{sec:calculations}.  We have found the following.

\begin {itemize}
\item[(i)] While the mass distribution of HFMWDs is not very sensitive
  to $\alpha$, good agreement can be obtained with both the observed
  mass distribution and the observed incidence of magnetism for models
  with {$0.1\,\le\,\alpha\,\le\,0.3$}.  In particular the mean
  predicted mass of HFMWDs is $0.88\,\rm\msun$ compared with
  $0.64\,\rm\msun$ (corrected to include observational biases) for all
  white dwarfs while observations indicate respective mean masses of
  $0.85\,\rm\msun$ \citep[see also][]{Kepler2013} and $0.62\,\rm\msun$
  \citep{Kleinman2013}.  A K--S test shows that the small number of
  reliably measured masses of HFMWDs are not distributed in the same
  way as the masses of non-magnetic single white dwarfs.  The
  probability they are is only $3\times 10^{-5}$.  On the other hand
  our best model fit to the observed mass distribution of HFMWDs has a
  probability of 0.71.
  
\item[(ii)] Stars that merge during common envelope evolution and then
  evolve to become white dwarfs outnumber merging post-common envelope
  DD systems for all $\alpha$.  The common envelopes yield mainly
  CO~white dwarfs with a few He and ONe~white dwarfs, while the DDs
  yield only CO~white dwarfs.

\item[(iii)] The major contribution to the observed population of
  HFMWDs comes from main-sequence stars merging with degenerate cores
  at the end of common envelope evolution.  The resulting giants go on
  to evolve to HFMWDs.

\item[(iv)]The merging DDs tend mostly to populate the high-mass end of
  the white dwarf mass distribution.

\end{itemize}
We also note that the study by \citet{Zorotovic2010} of the evolution
of a sample of SDSS post-common envelope binary
stars consisting of a white dwarf and a main-sequence star indicates
that the best agreement with observational data is achieved when
$\alpha=0.25$ and this is consistent with our findings.

In summary, available observations of the mass distribution and
incidence of HFMWDs are compatible with the hypothesis that they arise
from stars that merge mostly during common envelope evolution with a
few that merge during post-common envelope evolution as DD systems.
Our calculations, when taken together with the observation that there
are no examples of HFMWDs in wide binary systems, allow us to argue
strongly in favour of this hypothesis.  In the majority of cases the
fields may be generated by a dynamo mechanism of the type recently
proposed by \citet{wickramasinghe2014}.  The disc field model of
\citet{Nordhaus2011} or the model proposed by \citet{garcia2012} may
be relevant in the case of merging DD cores depending on mass ratio.
The rate of merging of post-common envelope DDs alone is too low to
account for all observed HFMWDs.

\section*{Acknowledgements}

We would like to thank the Referee, Zhanwen Han, for his suggestions
and comments which have helped us improving the quality of this
paper. GPB gratefully acknowledges receipt of an Australian
Postgraduate Award.  CAT thanks the Australian National University for
supporting a visit as a Research Visitor of its Mathematical Sciences
Institute, Monash University for support
as a Kevin Watford distinguished visitor and Churchill College for his fellowship.\\

\label{lastpage}

\end{document}